\documentclass[12pt, draftclsnofoot, onecolumn,romanappendices]{IEEEtran}
\usepackage{bm,cite}
\usepackage{amsmath}
\usepackage{amsthm}
\usepackage{amsbsy}
\usepackage{epsfig}
\usepackage{latexsym}
\usepackage{amssymb}
\usepackage{wasysym}
\usepackage{placeins}
\usepackage{subfig} 
\usepackage{graphicx}
\usepackage{multicol}
\usepackage{here}

\DeclareMathAlphabet{\mathbit}{OML}{cmr}{bx}{it}
\DeclareMathAlphabet{\mathsf}{OT1}{cmss}{m}{n}
\DeclareMathAlphabet{\mathTXf}{OT1}{cmss}{bx}{it}

\DeclareMathOperator{\Dist}{Dist}

\newcommand{\trans}{{\text{T}}} 
 
\newcommand{\He}{{{\text{H}}}}

\newcommand{\SNR}{{\text{SNR}}}  
\newcommand{\INR}{{\text{INR}}}

\begin{document} 
\title{CSI Sharing Strategies for Transmitter Cooperation in Wireless Networks}
\author{Paul de Kerret and David Gesbert\\Mobile Communications Department, Eurecom\\
Campus SophiaTech, 450 Route des Chappes, 06410 Biot, France\\\{dekerret,gesbert\}@eurecom.fr}

\maketitle
\begin{abstract}
Multiple-antenna ``based" transmitter (TX) cooperation has been established as a promising tool towards avoiding, aligning, or shaping the interference resulting from aggressive spectral reuse. The price paid in the form of feedback and exchanging channel state information (CSI) between cooperating devices in most existing methods is often underestimated however. In reality, feedback and information overhead threatens the practicality and scalability of TX cooperation approaches in dense networks. Hereby we addresses a ``Who needs to know what?" problem, when it comes to CSI at cooperating transmitters. A comprehensive answer to this question remains beyond our reach and the scope of this paper. Nevertheless, recent results in this area suggest that CSI overhead can be contained for even large networks provided the allocation of feedback to TXs is made non-uniform and to properly depend on the network's topology. This paper provides a few hints toward solving the problem.  
\end{abstract}

\section{Introduction}
Wireless communication has become essential to our lives in many ways, through a variety of services as well of devices ranging from pocket phones to laptops, tablets, sensors and controllers. The advent of multimedia dominated traffic poses extra-ordinary constraints on data rates, latency and above all spectral efficiency. In order to deal with the expected saturation of available resources in currently used bands, new wireless systems are designed based on $i)$ greater densification of infrastructure equipments (small cells), and $ii)$ a very aggressive spatial frequency reuse, which in turn results in severe \emph{interference} conditions for cell-edge terminals.  

The role played by multiple antenna combining in mitigating interference by means of zero forcing (ZF) (or related criteria) beamforming is well established. Over the last few years, the combination of multiple-antenna approaches together with the concept of cooperation among interfering wireless devices was explored, showing strong promise (see \cite{Gesbert2010} and references therein). In particular, TX-based cooperation allows for avoidance of the interference before it even takes place (e.g. multi-cell MIMO or ``Joint Processing CoMP"), or helps to shape it in a way which makes it easier for the receivers (RX) to suppress it (e.g. alignment). TX cooperation methods can be categorized depending on whether the data messages intended at the users must be known at several TXs simultaneously or not. For systems not allowing such an exchange (e.g. due to privacy regulations or low backhaul capabilities), interference alignment (IA) has been shown to be instrumental \cite{Cadambe2008}. In contrast, when user data message exchange is made possible by a specific backhaul routing architecture, multi-cell, a.k.a. ``network" MIMO, methods offer the best theoretical benefits \cite{Karakayali2006}. 

A distinct advantage of TX cooperation over conventional approaches relying on egoistic interference rejection, lies in the reduced number of antennas needed at each RX to ZF residual interference. This gain is further amplified when user data messages exchange among TXs is made possible. For instance, in the case of three interfering two-antenna TXs, relying on RX based interference rejection alone requires three antennas to ZF the interference at each RX, while just two are needed when coordination is enabled via IA\cite{Yetis2010}. Further, if the three user messages are exchanged among the TXs, thus enabling network MIMO precoding, then just one antenna per TX and RX is sufficient to preserve interference-free transmission. 

However, the benefits of multiple antenna transmit cooperation go at the expense of requiring channel state information (CSI) at the TXs. Indeed, whether one considers cooperation with or without user's data sharing, the TXs should in principle acquire the complete CSI pertaining to every TX and RX pair in the network. This is also the case for distributed schemes (e.g. \cite{Gomadam2011}) where  the computation of precoders typically relies on iterative techniques where each iteration involve the acquisition of local feedback. Yet, as local feedback is updated over the iterations, this approach implicitly allows each TX to collect information about the precoders and channels of other TXs, hence amounting to an iterative global CSI acquisition at all TXs. 

At first glance, CSI feedback and sharing requirements grow unbounded with the network size. Since over the air feedback and backhaul exchange links are always rate and latency limited, this means the practical application of TX cooperation in dense large networks is difficult. 

In this paper, we challenge the common view that interfering TXs engaging in a cooperative scheme \emph{can} or \emph{should} share global (network-wide) CSI. Instead, we formulate the problem of a suitable CSI \emph{dissemination} (or allocation) policy  across transmitting devices while maintaining performance close to the full CSIT sharing scenario. We report a couple of findings revealing how the need for CSIT sharing can be alleviated by exploiting specific antenna configurations or decay property of signal strength versus distance, hence making TX cooperation distributed and scalable. We use interference alignment and network MIMO respectively as our driving scenarios.

More specifically, for the cooperation scenario without user data message sharing where alignment of interference is sought, we show how perfect alignment is possible in certain antenna topologies without knowledge of all the channel elements at some TXs. For the network MIMO scenario, this is not the case and we illustrate instead how power decay versus distance can be exploited to substantially reduce the CSI sharing requirements while fulfilling optimal asymptotic rate performance conditions. A common trait behind the findings is that different cooperating TXs can (and often must) live with their own individual partial version of the global CSIT. Hence, CSIT representation quality is bound to be non-uniform across TXs. Consequently, we discuss briefly the problem of multiple-antenna precoding with \emph{TX-dependent} CSI.  

\section{Brief Notions in the Multiple-Antenna Transmitter Cooperation} 

We consider fast fading multiple-antenna wireless networks where the transmission can be mathematically represented by writing~$\bm{\mathrm{y}}_i$, the received signal at RX~$i$, as
\begin{equation}
\bm{\mathrm{y}}_i=\mathbf{H}_{ii}\bm{\mathrm{x}}_i+\sum_{j\neq i} \mathbf{H}_{ij}\bm{\mathrm{x}}_j
\end{equation}
where~$\bm{\mathrm{x}}_i$ is the signal emitted by TX~$i$ and~$\mathbf{H}_{ij}$ is a matrix containing the channel elements between TX~$j$ and RX~$i$. The transmitted symbols~$\bm{\mathrm{x}}=[\bm{\mathrm{x}}_1,\ldots,\bm{\mathrm{x}}_K]^{\trans}$ are then obtained from the user's data symbols~$\bm{s}=[s_1,\ldots,s_K]^{\trans}$ by multiplication with a precoder~$\mathbf{T}$, i.e., $\bm{\mathrm{x}}=\mathbf{T}\bm{s}$. If the user's data symbols are not shared between the TXs, the precoder~$\mathbf{T}$ is restricted to a particular block-diagonal structure, while it can otherwise take any form. The received filter $\bm{g}_i^{\He}$ is then applied to the received signal~$\bm{y}_i$ to obtain an estimate of the transmitted data symbol.

Here, we briefly discuss the leading techniques for MIMO based cooperation with or without user data message exchange. We point out commonly made assumptions in terms of CSIT sharing and feedback design. 

\subsection{Interference Alignment for Interference Channels}
When the user's data symbols are not shared between the TXs, the setting is referred to as a \emph{Interference Channel (IC)} in the communication theoretic literature. In MIMO ICs, a method called \emph{Interference Alignment (IA)} has been recently developed and shown to achieve the maximal number of degrees-of-freedom (DoF), or pre-log factor, in many cases\cite{Cadambe2008,Yetis2010}. As a consequence, IA has attracted a lot of interest in the community. In this work, we will take the DoF as our key performance metric, such that we focus on IA schemes.

IA is said to be \emph{feasible} if the antenna configuration (i.e. the distribution of antenna elements at the TXs and the RXs) yields enough optimization variables to allow for the interference-free transmission of all user's data symbols, which means fulfilling \cite{Yetis2010}
\begin{equation}
\forall i, \forall j\neq i, ~~\bm{g}_i^{\He}\mathbf{H}_{ij}\bm{t}_j=0.
\end{equation}
Intuitively, IA consists in letting the TXs coordinate among themselves to beamform their signals such that the interferences received at each of the RXs are confined in a subspace of reduced dimensions, which can then be suppressed by linear filtering at the RXs with a smaller number of antennas.
\subsection{Precoding in the Network MIMO} 

When the user's data symbols are shared between the TXs, the TXs form a \emph{distributed antenna array} and a joint precoder can be applied at the transmit side \cite{Karakayali2006}. Consequently, this setting becomes similar to the single TX multi-user MIMO downlink channel and the interference between the TXs can be completely canceled, e.g., by applying a global ZF precoder~$\mathbf{T}\propto\mathbf{H}^{-1}$.

\begin{figure}[ht!]
\centering
\subfloat[][]{
\label{IA}
\includegraphics[width=0.9\columnwidth]{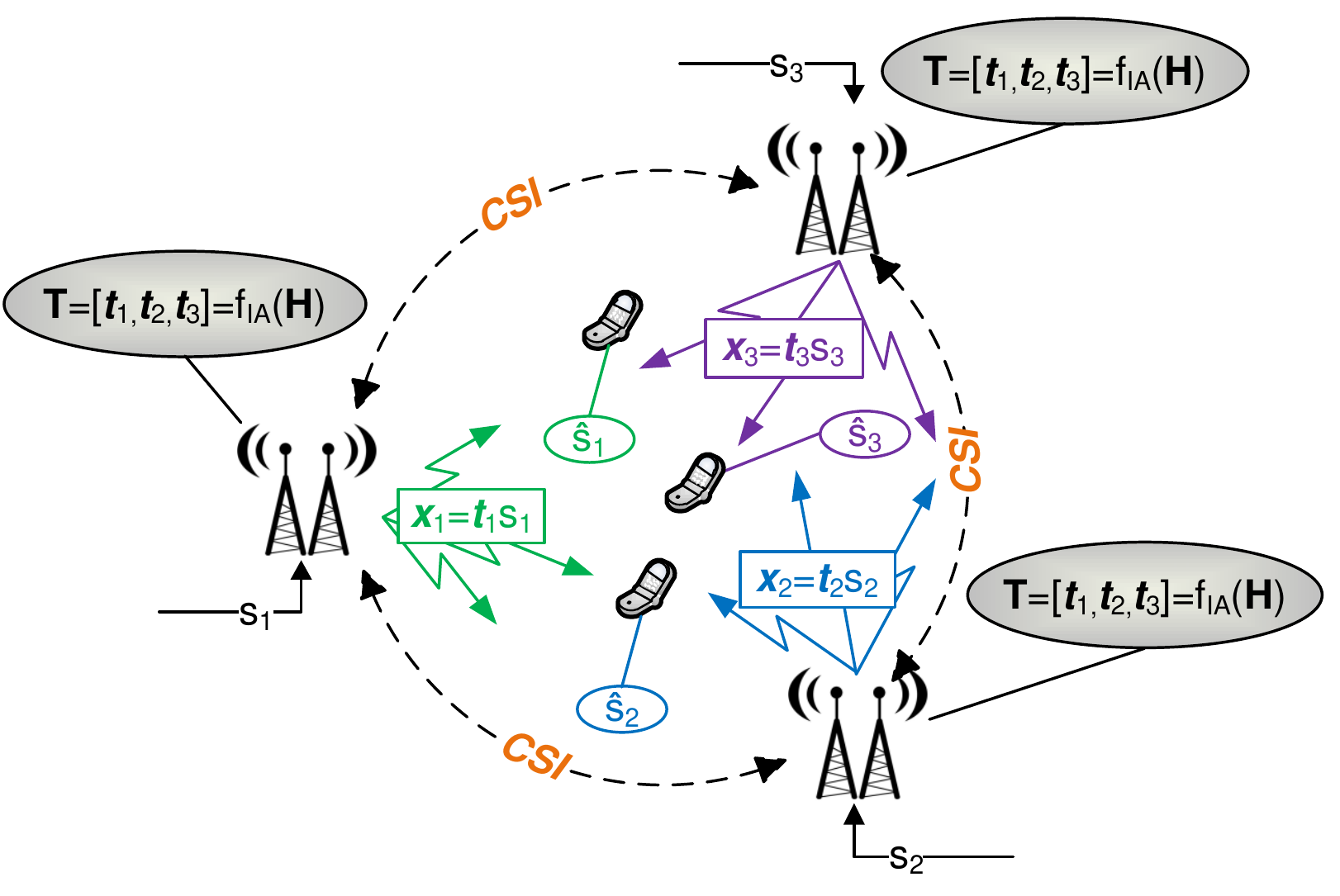}}
\hspace{8pt}  
\subfloat[][]{\label{JP}
\includegraphics[width=0.9\columnwidth]{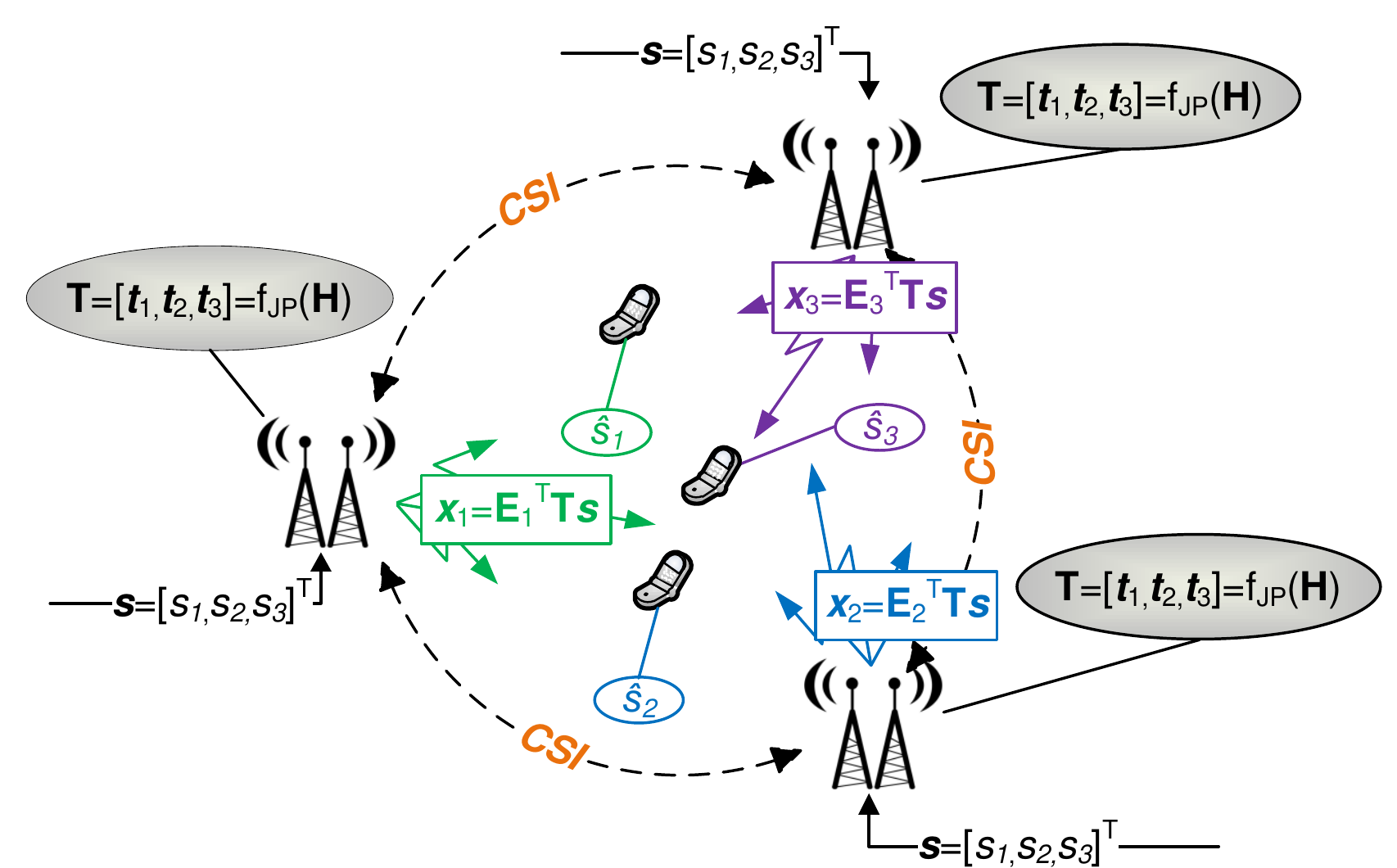}}
\caption[]{The description of a distributed IA algorithm is done in Figure~\subref{IA} while Figure~\subref{JP} represents the distributed precoding in a Network MIMO with user's data sharing. The matrix~$\mathbf{E}_i$ is a matrix which selects the rows of the multi-user precoder~$\mathbf{T}$ corresponding to the antennas at TX~$i$.} 
\label{IA-JP}%
\end{figure}

These two scenarios are schematically represented in Figure~\ref{IA-JP}.


\subsection{Limited Feedback Versus Limited Sharing}\label{se:Feedback}
The limited feedback capabilities have been recognized as a major obstacle for the practical use of the precoding schemes described above. Consequently, a large literature has focused on this problem and both efficient feedback schemes and robust transmission schemes have been derived, for Network MIMO \cite{Shenouda2006,Jindal2006} and IA \cite{Bolcskei2009,ElAyach2012}.

Yet, all these works assume that the imperfect channel estimates obtained via limited feedback are \emph{perfectly} shared between all the transmit antennas. This is a meaningful assumption when the TXs are colocated but less realistic otherwise, as we shall now see. 

\subsubsection*{CSIT Sharing Issues}\label{se:Feedback}
One obstacle to the sharing of global CSIT follows from the fact that the amount of CSI which has to be exchanged increases very quickly with the number of TXs. In fact, each TX needs to obtain the CSI relative to the full multi-user channel, which consists of $(NK)^2$ scalars in a $K$-user setting with $N$ antennas at each node.

\begin{figure}[ht!] 
\includegraphics[width=1\columnwidth]{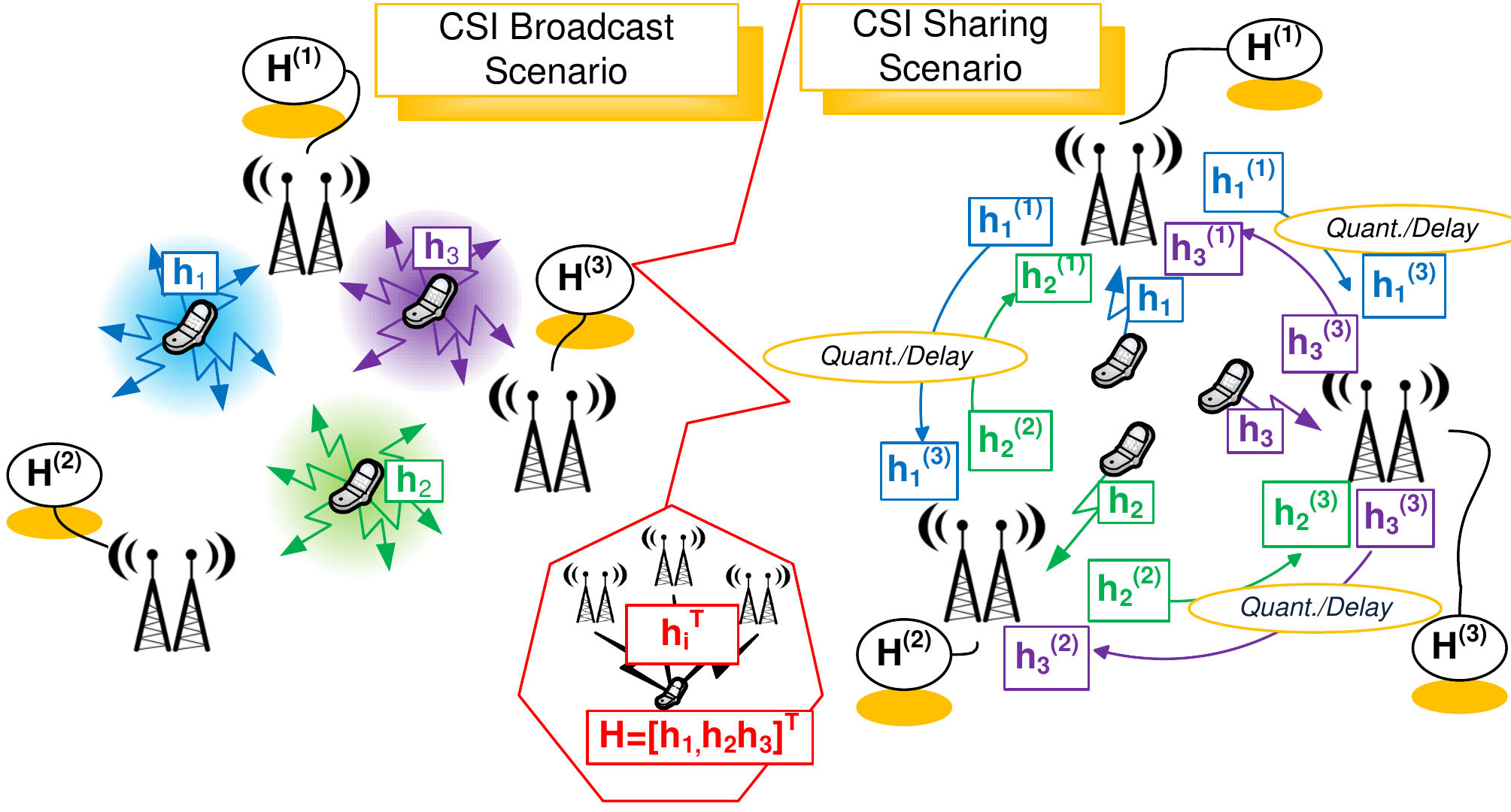} 
\caption[]{Possible scenarios for the feedback of the CSI.} 
\label{CSIT_sharing}%
\end{figure}

In addition, acquiring the CSI at a particular TX can be realized either by a direct broadcast of the CSI to all the listening TXs or by an over-the-air feedback to the \emph{home} base station alone, followed by an exchange of the local CSIs over the backhaul, as it is currently advocated by 3GPP LTE-A standards\cite{Sesia2011}. See the illustrations for such scenarios in Figure~\ref{CSIT_sharing}. Note that exchange over the backhaul can involve further quantization loss and may lead to a different CSI-aging at each TX, due to protocol latency. Either case, the channel estimates available at the various TXs will not be exactly the same. This leads to a form of CSI discrepancy which is inherent to the cooperation among non-colocated TXs.

In order to capture multiple-antenna precoding scenarios whereby different TXs obtain an imperfect \emph{and} imperfectly shared estimate of the overall multi-user channel, we denote by $\mathbf{H}^{(j)}$ the network-wide channel matrix estimate available at TX~$j$. Consequently, the precoding schemes in Figure~\ref{IA-JP} have to be modified to take into account that each TX will compute its precoder based on its own channel estimate. Thus, TX~$j$ transmits~$\bm{\mathrm{x}}_j=\mathbf{T}^{(j)}\bm{s}$ based on the knowledge of $\mathbf{H}^{(j)}$ only. 

The fundamental questions which arise are: $(i)$ how complete and accurate should the estimate $\mathbf{H}^{(j)}$ be for each $j$ while operating under reasonable CSI overhead constraints? and $(ii)$ how should precoders be designed given the likely discrepancies between various channel estimates? Although these questions prove to be difficult and to a large extent remain open, we shed some light on the problem for two key scenarios in the following sections.

\section{Aligning Interference with Incomplete CSIT}\label{se:IA}

Let us first consider an IC, i.e., without user's data sharing. Feasibility studies for IA are typically carried out under the assumption of full CSIT. Yet, one can show that IA feasibility and the CSIT model are in fact tightly coupled notions. Assume for instance that all the RXs were given a generous number of antennas equaling or exceeding the number of TXs, it is well known that the interference could be suppressed at the RXs alone and no precoding, and hence no CSIT, is necessary. This example suggests the existence of a trade-off between the number of antennas and the CSI sharing requirements. Thus, it is possible to design IA algorithms using less CSIT that conventionally thought, without performance degradations by exploiting the availability of extra-antennas at a subset of devices. More specifically, the problem of finding the minimal CSIT allocation which preserves IA feasibility can be formulated. The minimality refers to the size of a CSIT allocation, defined as the total number of scalars sent through the multi-user feedback channel. 

We differentiate between antenna configurations where IA is feasible and the number of antennas at the TXs and the RXs provide just enough optimization variables to satisfy alignment conditions, denoted as \emph{tightly-feasible}, and the ones where extra antennas are available, denoted as \emph{super-feasible}. Furthermore, we call a CSIT allocation \emph{strictly incomplete} if at least one TX does not have the complete multi-user CSI. With such concepts in place, the following lesson can be drawn.

\subsection{Tightly-feasible ICs}
A strictly incomplete CSIT allocation implies that some TXs compute their precoders in order to fulfill IA inside a smaller IC formed by a subset of RXs and a subset of TXs. Most of the time, this creates additional constraints for the optimization of the other precoders which makes IA unfeasible. Yet, it can be shown that IA feasibility can be preserved under the following condition \cite{dekerret2012_ISWCS_journal}.

\newtheorem{lesson}{Lesson}
\begin{lesson}
In a tightly-feasible IC, there exists a strictly incomplete CSIT allocation preserving IA feasibility if there exists a tightly-feasible sub-IC strictly included in the full IC. 
\end{lesson}
Exploiting this result, a CSIT allocation algorithm is derived in~\cite{dekerret2012_ISWCS_journal} along with an algorithm which achieves IA based on this incomplete CSIT allocation. In a few words, it consists in giving to each TX the CSI relative to the smallest tightly-feasible sub-IC to which it belongs. The precoders are then designed to align interference inside this smaller sub-IC, thence requiring only a part of the total CSIT, while IA feasibility is preserved. We will see in the simulations results presented in the following that the reduction in the CSIT size is significant. In fact, the reduction of the CSIT allocation feeds on the heterogeneity of the antenna configuration such that the more heterogeneous the antenna configuration is, the larger is the saving brought by using the minimal CSIT allocation. This is particularly appealing in regards to the future networks where mobile units and base-stations from different generations with different number of antennas are likely to co-exist.

\subsection{Super-feasible ICs}

In super-feasible settings, the additional antennas can be used to reduce the size of the minimal CSIT allocation. Yet, how to exploit optimally these additional antennas to reduce the feedback size is a very intricate problem. Still, a low complexity heuristic CSIT allocation can be derived \cite{dekerret2012_ISWCS_journal}. The main idea behind the algorithm is to let some TXs or RXs ZF less interference dimensions such that small tightly-feasible settings are formed inside the original setting. 

\begin{figure}
\centering
\includegraphics[width=1\columnwidth]{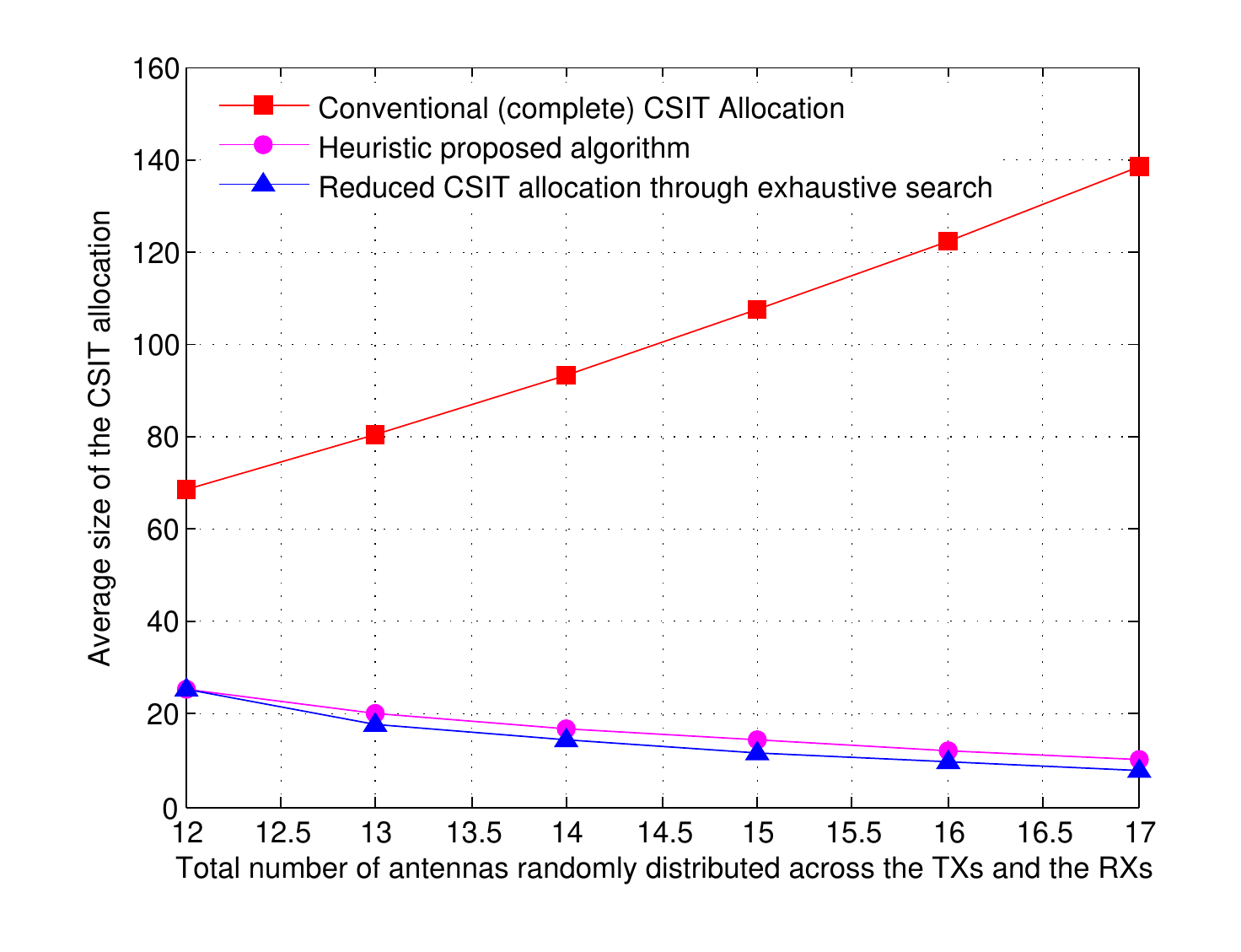}
\caption{Average CSIT allocation size in terms of the number of antennas randomly distributed across the TXs and the RXs for $K=3$~users.}
\label{Feedback_size_K3}
\end{figure}

The effective CSIT reduction is illustrated in Figure~\ref{Feedback_size_K3} for a $3$-user IC. The results are averaged over $1000$ random distributions of the antennas across the TXs and the RXs. If $12$~antennas are distributed between the TXs and the RXs, the setting is tightly-feasible and the previous CSIT allocation policy for tightly-feasible settings is used. With more than $12$~antennas, the algorithm exploits every additional antenna to reduce the size of the CSIT allocation.

When the setting is tightly-feasible, the reduction in feedback size requires neither a DoF reduction nor any additional antenna and comes in fact ``for free": It simply results from exploiting the heterogeneity in the antenna numbers at the TXs and the RXs.
\section{A CSIT Allocation Policy for Network MIMO}\label{se:Distance}
When the user's data symbols are jointly precoded at the TXs, complete CSIT allocation, in the sense defined above, is needed in all the practically relevant scenarios. So in this case, a different notion of reduced CSIT sharing must be advocated. The essential ingredient of this approach is the classical intuition that a TX should have a more accurate estimate for channels creating the strongest interference, i.e. originating from devices in the close neighborhood. This means that the fact that interference decays with pathloss can be exploited in  principle to reduce the CSIT sharing requirements. This concept  was  recently introduced in \cite{dekerret2012_ICC}. A mathematical tool known as generalized degree-of-freedom comes handy to capture the effect of path loss on the multiplexing gain of cooperating networks with partially shared CSIT. Additionally, a simplified model referred to as Wyner model is used in this context to aid analytical tractability and provide first insights into this problem. 

\subsection{Generalized Degrees of Freedom}
TX cooperation methods are often evaluated through the prism of DoF performance. Unfortunately the DoF is essentially pathloss-independent, such that a DoF analysis fails to properly capture the behavior of a large (extended) network MIMO. An extension of the notion of DoF, introduced in~\cite{Etkin2008} as the \emph{generalized DoF}, offers a much better grip over the problem as it can better take pathloss models into account. Upon defining the \emph{interference level}~$\gamma$ as~$\gamma\triangleq \log(\INR)/\log(\SNR)$ with $\SNR$ denoting the signal-to-noise ratio and $\INR$ the interference-to-noise ratio, it is possible to define the \emph{generalized DoF} as the DoF obtained when the SNR and the INR tend both to infinity for {\em a given interference level~$\gamma$}. 

For ease of exposition, we consider scenarios where all the TXs and RXs have a single antennas. The CSI is distributed, meaning that each TX has its own channel estimate based on which it computes its transmit coefficient without further exchange of information with the other TXs. We denote the estimate at TX~$j$ by $\mathbf{H}^{(j)}$ and its $i$-th row, which corresponds to the channel from all TXs to RX~$i$, by $\bm{h}_i^{(j)}$. We consider a digital quantization with a number of bits quantization~$\bm{h}_i^{(j)}$ equal to~$B_i^{(j)}$. Therefore, TX~$j$ computes its own version of the precoding matrix~$\mathbf{T}^{(j)}$ based on its own estimate~$\mathbf{H}^{(j)}$. It then transmits~$\mathrm{x}_j=\bm{e}_j^{\trans}\mathbf{T}^{(j)}\bm{s}$.

\subsection{The One Dimensional Wyner Model}
\begin{figure}
\centering
\includegraphics[width=1\columnwidth]{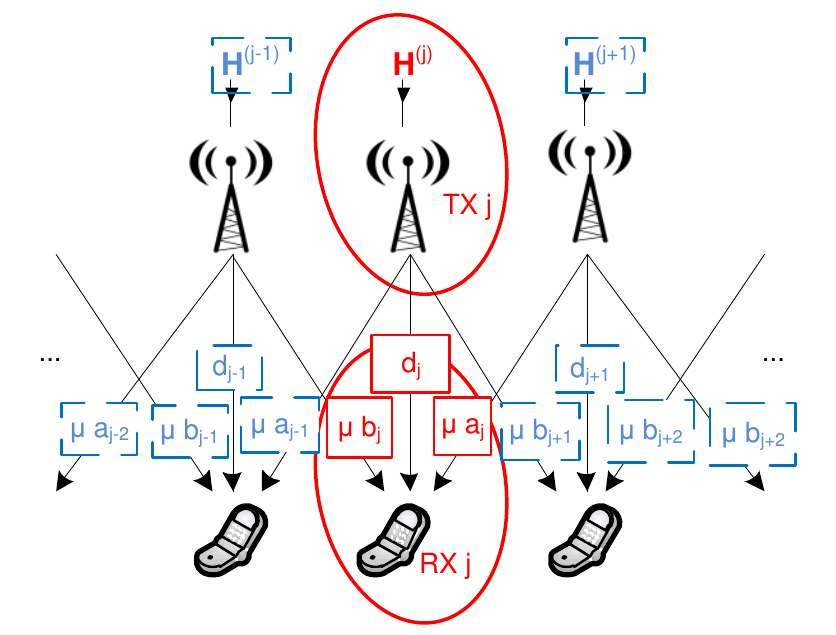}
\caption{Schematic representation for the Wyner model considered.}
\label{System_Model_Wyner}
\end{figure}

In the simple 1D \emph{Wyner model}\cite{Wyner1994}, the TXs are regularly placed along a line and solely the direct neighboring TXs emit non-zero interference. The channel is thus represented by a tridiagonal matrix. Furthermore, we assume that the interference from the direct neighbors are attenuated with a coefficient~$\mu=P^{\gamma-1}$, according to the generalized DoF model. The transmission in the Wyner model is schematically presented in Figure~\ref{System_Model_Wyner}.


Our objective is to evaluate how small $B_i^{(j)}$ can be while guaranteeing the same DoF as a system with perfect CSIT. Obviously, sharing the most accurate CSIT to all the TXs is a possible solution, yet, the size of the CSI required \emph{at each TX} grows then unbounded with the number of users~$K$, making this solution both inefficient and unpractical. 
In contrast, a much more efficient CSI sharing policy achieving the maximal generalized DoF, denoted as \emph{distance-based}, is summarized below \cite{dekerret2012_ICC}.

\subsection{Distance-Based CSIT Allocation}
We are interested in a CSIT allocation strategy, referred below as ``distance-based", whereby each TX receives a number of CSI scalars which remains \emph{bounded} as the number of users~$K$ increases.

The distance-based CSIT allocation is obtained by setting for all $i,j$, 
\begin{equation}  
B_i^{(j)\Dist}\!=\!\lceil([1\!+\!(\gamma\!-\!1)|i\!-\!j|]^{+}+2[\gamma+(\gamma\!-\!1)|i\!-\!j|]^{+})\log_2(P)\rceil
\label{eq:eq_3}
\end{equation}
where $[\bullet]^{+}$ is equal to zero if the argument is negative and to the identity function otherwise, and $\lceil \bullet \rceil$ is the ceiling operator. It can be shown that the CSIT allocation~$\{B_i^{(j)\Dist}\}_{i,j}$ allows to achieve the maximal generalized DoF, i.e., those achieved in a system with perfect feedback \cite{dekerret2012_ICC}. The proof is based on the off-diagonal exponential decay of the inverse of the tridiagonal channel matrix.

Setting~$\gamma=1$ (no significant pathloss attenuation) in the previous equation, a conventional CSIT allocation is obtained where all channels are described with the same number of bits at all TXs (uniform allocation). For $\gamma<1$ however, the number of bits allocated decreases with the distance~$|i\!-\!j|$ between the considered TX and the index of the quantized channel, until no bit at all is used for quantizing the channel if the distance $|i-j|$ between the RX and the TX is larger than $\lceil 1/(1-\gamma)\rceil$. Crucially, this solution allocates to each TX a total number of bits which no longer grows with~$K$ as only the CSI relative to a \emph{neighborhood} is shared at each TX.

\begin{figure}
\centering
\includegraphics[width=1\columnwidth]{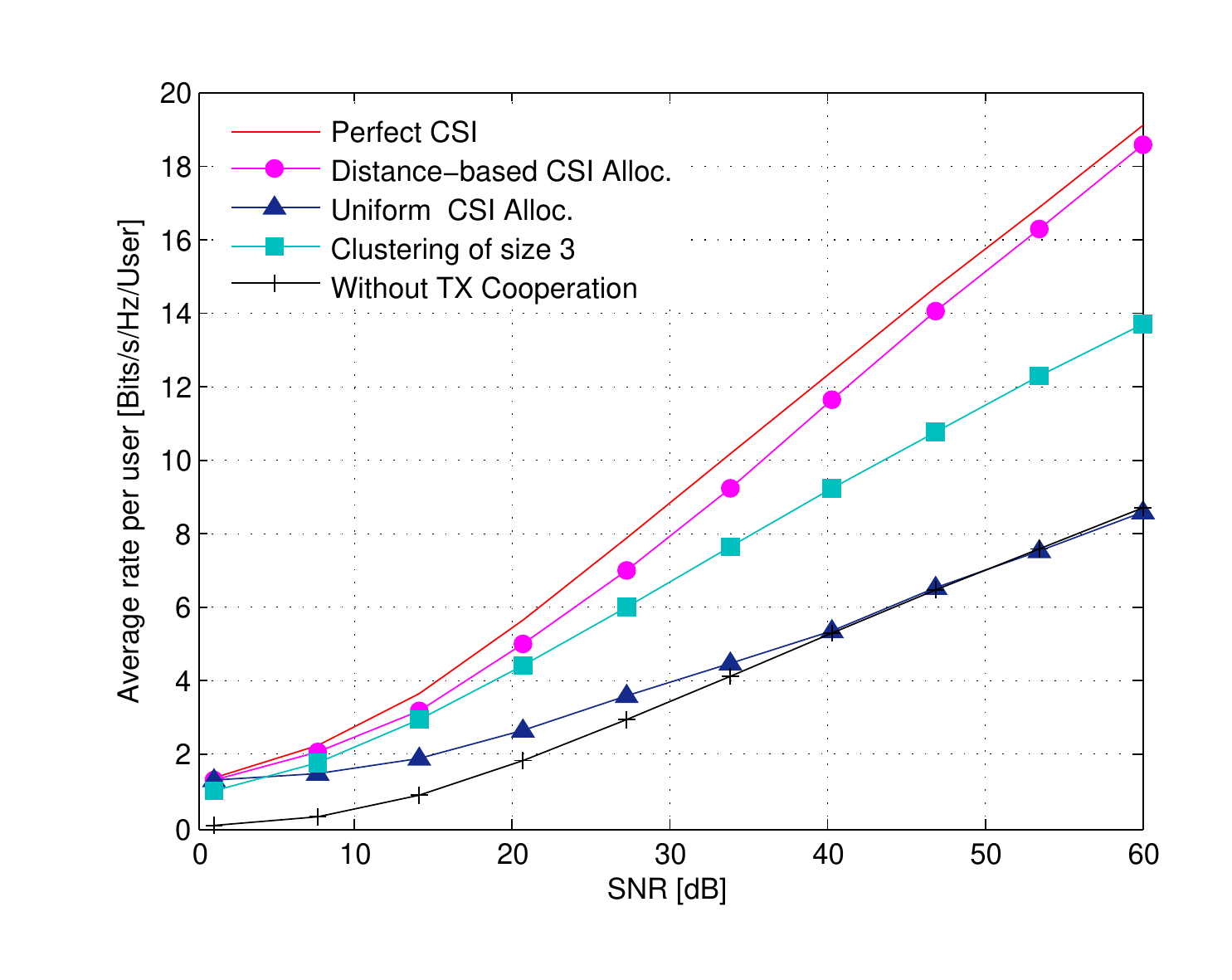} 
\caption{Average rate per user in terms of the SNR. The distance-based CSIT allocation, the uniform allocation, and the clustering one have all a size equal to $10\%$ of the size of the conventional CSIT allocation.}
\label{Fig_Dist}%
\end{figure}

The different CSIT allocations are compared in Figure~\ref{Fig_Dist} for $K=15$~users and~$\gamma=0.5$. The conventional CSIT allocation consists in providing the best quality to all the TXs, while the other strategies have the same size as the distance-based CSIT allocation, but the feedback bits are respectively shared uniformly and according to a conventional clustering of size~$3$. It can be seen from~\eqref{eq:eq_3} that the ratio between the size of the distance-based CSIT allocation and the conventional CSIT allocation is independent of the SNR. Here, the distance-based CSIT allocation represents only~$10\%$ of the size of the conventional CSIT allocation. 

Hence, the distance-based CSIT allocation achieves the maximal number of generalized DoF with only a small share of the total CSIT, and outperforms the other schemes of comparison. Additionally, the user's data sharing can also be reduced to a neighborhood without loss of performance. Consequently, this scheme can be seen as an alternative to clustering in which the hard boundaries of the clusters are replaced by a smooth decrease of the level of cooperation. 

\section{Precoding in the Network MIMO Channel with Distributed CSIT}\label{se:DCSI}

As it becomes clear from the previous section, an efficient CSI dissemination policy naturally leads to a significant reduction of the CSI sharing requirements. As a consequence, the CSIT is represented non-uniformly across the TXs. Since non-uniform sharing is the best strategy in order to maximize performance under a given feedback overhead constraint, it is a natural consequence that some user's channels will be coarsely described at certain TXs and more accurately at others. Interestingly, the problem of designing precoders that can accommodate such a peculiar CSIT scenario is by and large open. In particular, new robust precoding schemes should be developed, as conventional precoders are designed under the assumption that the same imperfect CSIT is shared {\em perfectly} among the TXs. 

Let us consider the model of distributed CSI described in Section~\ref{se:Distance} where TX~$j$ receives its own channel estimate~$\mathbf{H}^{(j)}$ with the $i$-th row, denoted by~$\bm{h}_i^{(j)}$, obtained using $B_i^{(j)}$ quantizing bits. Assume a network where each TX has roughly the same average pathloss to each RX. The DoF which can be achieved with limited feedback is studied in \cite{Jindal2006} for the single TX MIMO downlink channel. We can extend this to the setting of non-uniform CSI so as to gain insight into the design of efficient precoders. In this case, \emph{CSI scaling} coefficients~$\alpha_i^{(j)}$ are introduced and defined by the limit of~$B_i^{(j)}/((K-1)\log_2(P))$ when $P$ goes to infinity. 

ZF is widely used and well known to achieve the maximal DoF in the MIMO downlink channel with perfect CSIT~\cite{Jindal2006}. One may wonder how conventional ZF performs in the presence of CSI discrepancies brought by imperfect sharing. The answer is strikingly pessimistic: The sum DoF achieved can be shown to be equal to just $K\min_{i,j} \alpha_i^{(j)}$\cite{dekerret2012_ISIT_journal}. Intuitively, this can be restated as follows.

\begin{lesson}
In the network MIMO with distributed CSI, the worst channel estimate across the TXs and the users limits the DoF achieved by each user using ZF precoding.
\end{lesson}
This is in strong contrast to the single TX case studied in \cite{Jindal2006} where the quality of the feedback of user~$i$ relative to~$\bm{h}_i$ solely impacts the DoF of user~$i$. It shows clearly the disastrous impact of the CSI non-uniformity, since one inaccurate estimate at one TX degrades the performances of \emph{all} the users. One may also wonder whether conventional-type robust precoders \cite{Shenouda2006} can offer a better response. The answer is negative, unfortunately. Instead, a novel precoder design is needed that is tailored to the non-uniform CSIT sharing model.

Preliminary results to this end \cite{dekerret2012_ISIT_journal} suggest that it is possible to dramatically improve the DoF in certain scenarios. For instance, in the two-TX network, a scheme referred to as \emph{Active-Passive (AP) ZF}, consisting in letting the TX with degraded CSIT arbitrarily fix its transmit coefficient while the other TX compensates to zero-out the interference, can be shown to recover the optimal DoF.

\begin{figure}
\centering
\includegraphics[width=1\columnwidth]{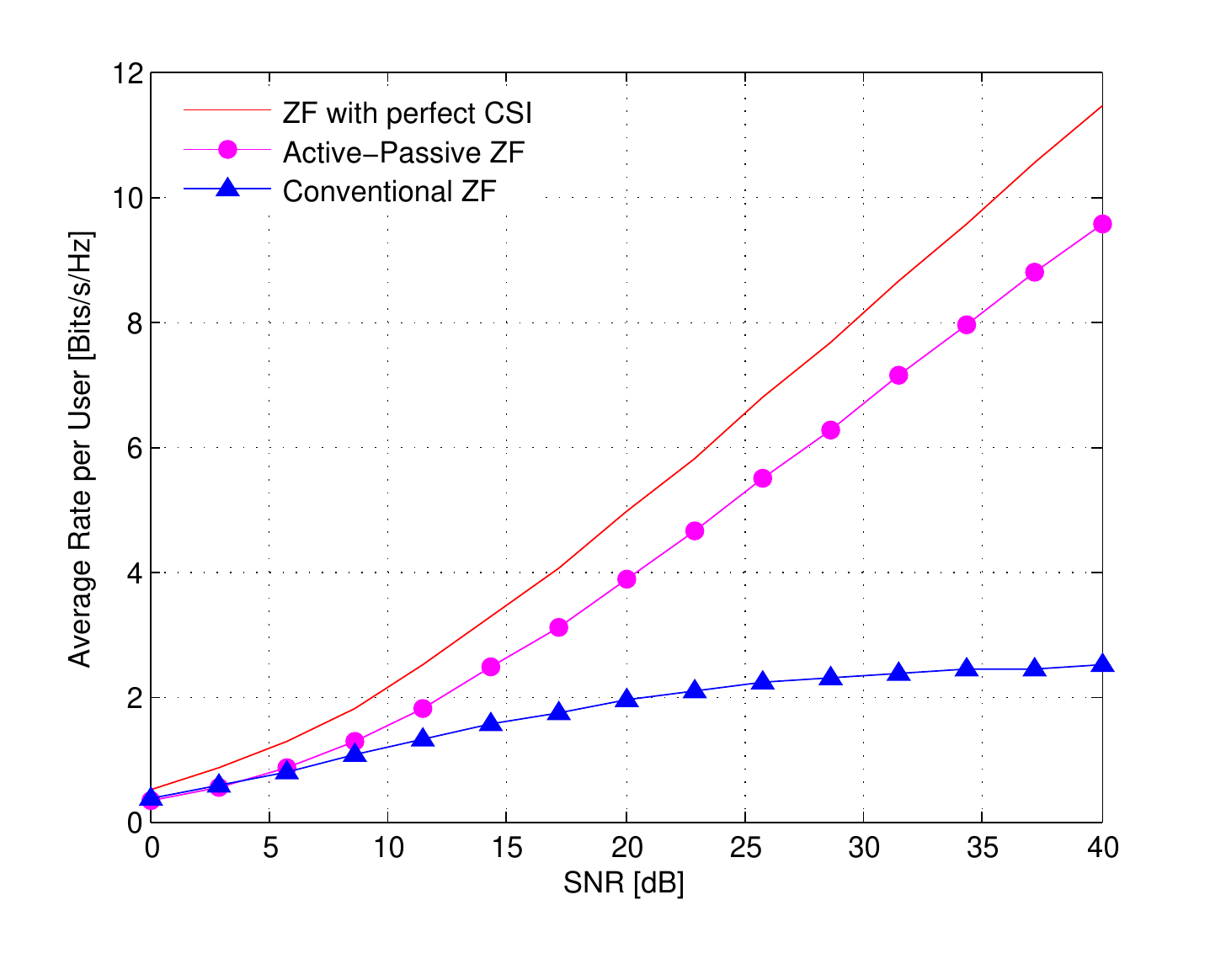}
\caption{Average rate per user in terms of the SNR for $\alpha_1^{(1)}=1$, $\alpha_2^{(1)}=0$, $\alpha_2^{(1)}=0.5$, $\alpha_2^{(2)}=0.7$.}
\label{Rate_1_05_0_07_stat}
\end{figure}

The average rate achieved with conventional ZF, AP ZF, and ZF with perfect CSIT are compared in Figure \ref{Rate_1_05_0_07_stat}. In that case, the sum rate of conventional ZF saturates at high SNR while AP ZF is more robust and achieves a better DoF.

\section{Open Problems}
New concepts for the CSIT sharing in wireless networks have been derived and their potential to reduce signaling overhead has been shown. We have presented some insights into a new problem which presents serious challenges, but also opportunities for the future. This leads to new intriguing open questions. Firstly, IA algorithms with incomplete CSIT are based on a DoF-preserving criterion only, i.e., on the performance at asymptotically high SNR. The impact of the incomplete CSIT on the performance at finite SNR should then be investigated to obtain practical solutions. Similarly, robust precoding schemes for the MIMO network with distributed CSI have so far considered DoF only as a metric. By and large, precoding over network MIMO with distributed CSI remains a challenging problem. Regarding the optimization of the CSIT allocation in a network MIMO channel, we believe that the distance-based CSIT allocation has the potential for impacting the design of cellular systems in practical settings. Yet, it has been examined for simplified channel models only and should be adapted to more realistic scenarios.

\section{Conclusion}  
The uniform allocation of CSI resources towards all TXs does not lead to an efficient use of the feedback and backhaul resources: The CSIT dissemination should be designed to allocate to each TX the right CSI, in terms of which elements to share and in terms of accuracy. For both network MIMO and IA, the CSIT sharing requirements have been significantly lowered by designing appropriate CSIT dissemination policies, thus making TX cooperation more practical and thereby paving the way for large performance improvement. Additionally, the problem of robust precoding schemes taking into account the inconsistency between the CSIs at the TXs has been tackled and shown to be a key element.

\section{Acknowledgment}
This work has been performed in the framework of the European research project HARP, which is partly funded by the European Union under its FP7 ICT Objective 1.1 - The Network of the Future.


\begin{biography}{Paul de Kerret} (IEEE Student Member) was born in 1987 in Paris, France. In 2009, he graduated from Ecole Nationale Superieure des Telecommunications de Bretagne, France and obtained a diploma degree in electrical engineering from Munich University of Technology (TUM), Germany. He also earned a four year degree in mathematics at the Universite de Bretagne Occidentale, France in 2008. From january 2010 to september 2010, he has been a research assistant at the Institute for Theoretical Information Technology, RWTH Aachen University, Germany. Since october 2010, he is working toward a Ph.D. degree in the Mobile Communications Department at EURECOM, France. 
\end{biography}

\begin{biography}{David Gesbert} (IEEE Fellow) is Professor and Head of the Mobile Communications 
Department, EURECOM, France. He received his Ph.D  in 1997 from Telecom Paris Tech. From 
1997 to 1999 he has been with the Information Systems Laboratory, Stanford University. In 
1999, he was a founding engineer of Iospan Wireless Inc, San Jose, Ca., a startup company 
pioneering MIMO-OFDM (now Intel). D. Gesbert has guest edited 6 special issues, published 
about 200 papers and several patents in the area wireless networks, among which three 
won international awards.
He co-authored the book “Space time wireless communications: From parameter estimation to 
MIMO systems”, Cambridge Press, 2006.
\end{biography}

\newpage

\newpage

\newpage

\newpage

\newpage

\newpage

\end{document}